\documentstyle[12pt]{article}
\newcommand{\gapproxeq}{\lower .7ex\hbox{$\;\stackrel{\textstyle >}{\sim}\;$}}
\newcommand{\lapproxeq}{\lower .7ex\hbox{$\;\stackrel{\textstyle <}{\sim}\;$}}
\newcommand{\eqn}[1]{(\ref{#1})}
\newcommand{\be}{\begin{equation}}
\newcommand{\ee}{\end{equation}}
\newcommand{\tnabla}{{\nabla}}
\newcommand{\bea}{\begin{eqnarray}}
\newcommand{\eea}{\end{eqnarray}}
\newcommand{\bean}{\begin{eqnarray*}}
\newcommand{\eean}{\end{eqnarray*}}

\newcommand{\I}{\mbox{\rm I} \hspace{-0.5em} \mbox{\rm I}\,}

\def\cstok#1{\leavevmode\thinspace\hbox{\vrule\vtop{\vbox{\hrule\kern1pt
\hbox{\vphantom{\tt/}\thinspace{\tt#1}\thinspace}}
\kern1pt\hrule}\vrule}\thinspace}

\begin{document}
\setlength{\unitlength}{1mm}
{\hfill Preprint DSF-T-1/95; DAMTP/95-15}

\hfill gr-qc/9503040\\
\begin{center}
{\Large\bf Inflationary Cosmology from Noncommutative Geometry }\\
\end{center}

\bigskip\bigskip

\begin{center}
{{\bf F. Lizzi$^1$}, {\bf G. Mangano$^{1,2}$}, {\bf G. Miele$^1$} and
{\bf G. Sparano$^1$}}
\end{center}

\begin{center}
{$^1$ {\it Dipartimento di Scienze Fisiche, Universit\`a di Napoli -
Federico II -, and INFN\\
Sezione di Napoli, Mostra D'Oltremare Pad. 20, 80125, Napoli, Italy}}\\
\bigskip
{$^2$ { \it Department of Applied Mathematics and Theoretical Physics, \\
Silver Street, CB3 9EW, Cambridge, United Kingdom}}
\end{center}

\bigskip\bigskip\bigskip

\begin{abstract}
In the framework of the Connes-Lott model based on
noncommutative geometry, the basic features of a gauge
theory in the presence of gravity are reviewed, in order to show the possible
physical relevance of this scheme for inflationary cosmology. These models
naturally contain at least two scalar fields, interacting with each other
whenever more than one fermion generation is assumed. In this paper
we propose to investigate the behaviour of these two fields (one of which
represents the distance between the copies of a two-sheeted space-time)
in the early stages of the universe evolution. In particular the
simplest abelian model, which preserves the main characteristics of more
complicate gauge theories, is considered and the corresponding inflationary
dynamics is studied. We find that a chaotic inflation is naturally favoured,
leading to a field configuration in which no symmetry breaking occurs and
the final distance between the two sheets of space-time is smaller the
greater the number of $e$-fold in each sheet.
\end{abstract}

\vspace{2.cm}
\centerline{{\it To be published in the International Journal of Modern
Physics A}}
\newpage
\baselineskip=.8cm

\section{Introduction}

The early universe is certainly the best laboratory to probe the structure of
space and time at very short distances, and during this period of the universe,
inflation plays a central role in solving important problems occurring in
standard cosmology. It is usually assumed that the dynamics of scalar fields is
responsible for inflation, by producing an effective cosmological constant term
in the Einstein equations. This leads, as well-known, to a quasi de Sitter
epoch during which the universe scale factor grows almost exponentially. The
knowledge of quantum field effects, in this era, is particularly relevant
because they are the origin of the subsequent structure of the present
universe. Thus, all models in which some scalar fields are present can be used
in principle to describe inflation, provided that the corresponding dynamics is
appropriate.

In this respect, the recently proposed gauge models inspired by noncommutative
geometry are a relevant example of theories in which scalar fields and
corresponding interaction potentials emerge naturally from geometry. In this
paper we study, for a wide class of these models, the predictions on the
resulting inflationary scenarios.

The programme of noncommutative geometry \cite{Book} shifts the stress {}from
space and time, traditionally seen  as a geometrical ensemble of points, to the
(commutative) algebra of continuous complex valued functions defined on it,
with the perspective of possible noncommutative generalizations. One of the
most interesting and promising products of this programme is the formulation of
the electroweak standard model by Connes and Lott \cite{ConnesLott,Varilly},
generalized to include Grand Unified Theories in \cite{Zurich}. Loosely
speaking, this formulation is based on a {\it doubling} of space--time,
considered as a two sheeted manifold. The algebra of continuous functions and
the Hilbert space of fields is thus doubled as well. The nontrivial feature of
the theory is that the Dirac operator, as well as the gauge potential
(connection), have some nondiagonal elements, which couple the two sheets of
space--time. These are classical scalar fields: one is related to the component
of the metric in the {\it discrete direction}, and thus to the distance between
the two sheets of space--time, and the others are Higgs fields, responsible for
the breaking of the symmetry. We will discuss in more detail the field content
of the theory in the first sections.

The action is as usual the square of the curvature of the connection. As the
Dirac operator contains the metric structure of a manifold, the presence of
this coupling gives a theory of gravity as well \cite{Zurgrav,gravity}. By
simply postulating that at the level of two-sheeted algebra the action is the
square of the curvature one obtains the Einstein-Hilbert Yang-Mills action
containing naturally an Higgs potential.

Interestingly, at the classical level this scheme generally predicts relations
between the physical masses of particles, such as the ratio of the top quark
mass over the Higgs mass, and fixes in some ranges the values of relevant
constants appearing in the electroweak standard model \cite{Kastler}.

The aims of such a programme are, in the long run, the prediction of physical
masses or other experimentally testable quantities, and even more the creation
of a unique framework to include gravity. In this initial stage, however, the
stress is more on the understanding of how the noncommutative structure of
space--time might account for the already known features of this (exceptionally
well working) electroweak model how we understand it presently. This
understanding, combined with the tools of noncommutative geometry, should then
provide the framework for the unveiling of deeper symmetries of the model, and
the possibilities to go further towards the unification with gravity.

At the moment all studies have been performed at the classical level, and
quantization can only be achieved with the usual field theory tools. This
quantization, however, although it does not respect the relations between the
coupling constants \cite{Madrid}, under the renormalization flow does not
change considerably these relations over various orders of magnitude of the
scale. A fully satisfactory achievement of the programme would be a
quantization made on the level of the algebra, and therefore intrinsically
related to noncommutative geometry. This remains one of the main challenges of
the programme.

On more phenomenological grounds, there is also a problem of scales which
requires further studies. The scalar field which is part of the Riemann
connection and is related to the distance between the two sheets of
space--time, pertaining to the intrinsic structure of the latter, has as a
natural scale, the Planck length. On the other side, according to the Connes
and Lott model \cite{ConnesLott}, at low energy this distance must be the
inverse of the top quark mass, which is several order of magnitude larger than
the Planck length, to correctly reproduce the electroweak standard model. We
would like to stress that an important feature of our work is to look at the
distance as undergoing a dynamical evolution. It remains a challenging
programme to study the  dynamics responsible for the subsequent evolution of
the distance down to the electroweak scale.

In this paper we study the scalar dynamics at the Planck energy since we are
interested in the cosmological implications of the noncommutative scenarios. As
we already noticed the natural epoch in which the scalar fields, related to the
connections in the discrete direction, may play a fundamental role is the
inflationary one. Therefore, we have studied, for the simplest abelian theory,
all the relevant characteristic which a {\it good} inflationary model should
satisfy: a quite large $e$-fold number $N$, large enough $(N \simeq 10^2)$ to
solve the flatness problem of standard cosmology; an efficient reheating stage,
which consists in the transfer of the enormous energy density stored in the
inflaton fields into entropy density for the relativistic degrees of freedom.

It is worth-while stressing that, although our analysis has been performed for
the simplest abelian model, the results are general enough to be looked upon as
indicating a typical behaviour of a noncommutative inspired gauge model. This
is because it already contains all the key features of more realistic dynamics
based on some non-abelian gauge symmetry group.

We cannot say in fact what the full gauge theory of fields and particles was at
epochs shortly after Planck time, and therefore we cannot decide the specific
model of noncommutative geometry to be used, namely which Hilbert space and
Dirac operator we must use. Nevertheless all considered models present at least
the fields mentioned above, namely two scalar fields which are the {\it
discrete components} of the gauge and gravitational connections.

This paper is organized as follows: in section 2 we review the model of
Yang-Mills theory plus gravity in noncommutative geometry, and obtain for the
abelian case the expression for the bosonic lagrangian. Section 3 is devoted to
the study of the scalar potential and to the related inflationary dynamics. The
one-loop effective potential is computed in the adiabatic approximation and the
results obtained solving the semiclassical scalar field equations of motion are
discussed. Finally we analyze, at one-loop level, the reheating mechanism by
introducing the coupling of the scalars to the fermionic degrees of freedom. It
is shown that this mechanism is efficient enough to make the transition from
the vacuum energy density to the radiation dominated universe. In section 4 we
give our conclusions and outlooks.

\section{General relativity and gauge theories in noncommutative geometry}

In this section the construction of a Einstein-Yang-Mills theory in
noncommutative geometry is briefly reviewed. The results, are then applied to
the abelian case. The reason of this choice is that, as far as the scalar
fields are concerned, the nonabelian aspect is practically irrelevant (apart of
the different irreducible representations to which the scalar fields belong to)
on the resulting inflationary dynamics. Thus, in this scenario, the abelian
theory not only represents a toy model for which it is interesting to study the
cosmological implications, but it also contains the essential ingredients of
any more complicate, but physically more relevant, gauge theory inspired by
noncommutative theory.

\subsection{Gravity}

In order to obtain the Einstein-Hilbert action \cite{Zurgrav} we first
introduce the notion of a $K$-cycle. A  $K$-cycle \cite{Book,Varilly,Zurich},
is a triple $({\cal A},{\cal H},D)$ where: ${\cal A}$ is a $*$-algebra
unitarily represented on the Hilbert space ${\cal H}$ in terms of bounded
operators, and $D$ is an unbounded selfadjoint operator with compact resolvent
such that the commutator with any element of the algebra is bounded. This
operator is the {\it generalized} Dirac operator. The usual concepts of
Riemannian geometry are now derived by the $K$-cycle. The general idea is the
following: the Riemannian metric is defined in terms of the $K$-cycle as a
hermitian structure $<.\, ,\, .>$ on $\Omega^1_D({\cal A})$, which is the space
of $1$-forms on ${\cal A}$ \cite{Book,Varilly} and thus is the noncommutative
analog of the cotangent bundle. As far as the notion of connection is
concerned, it is useful to think of a connection $\tnabla$ on a fiber bundle as
a linear map
\begin{equation}
\tnabla : E \longrightarrow \Lambda^1({\cal M})\otimes E \label{nabla}~~~,
\end{equation}
satisfying the Leibniz rule, where $E$ is the space of sections of the fiber
bundle and $\Lambda^1({\cal M})$ is the space of $1$-forms on the base manifold
${\cal M}$. In the case of a linear connection on ${\cal M}$, observing that we
are interested in forms rather than vector fields, the fiber bundle is taken to
be the cotangent bundle. Furthermore, if on ${\cal M}$ there is a metric $g^E$
(the index $E$ hereafter denotes the {\it Euclidean} version), a Levi-Civita
connection is a linear connection satisfying the requirements
\begin{eqnarray}
\tnabla g^E &=& 0 ~~~,\label{lc} \\
       T(\tnabla)\equiv d -
P \circ\tnabla  &=& 0~~~, \label{torsion}
\end{eqnarray}
where $T(\tnabla)$ is the torsion corresponding to the connection $\tnabla$,
and $P$ maps the tensor product of two $1$-forms into their exterior product.
The noncommutative notion of a Levi-Civita connection will then be given by
\eqn{nabla} with $E$ and $\Lambda^1$ both replaced by $\Omega^1_D({\cal A})$
\begin{equation}
\tnabla : \Omega^1_D({\cal A}) \longrightarrow \Omega^1_D({\cal A})\otimes
\Omega^1_D({\cal A})~~~.
\end{equation}
Of the two conditions \eqn{lc} and \eqn{torsion}, the former is naturally
implemented by requiring
\begin{equation}
d<\xi,\eta> = <\tnabla\xi,\eta> - <\xi,\tnabla\eta>~~~,
\end{equation}
whereas the latter can be straightforwardly used. The action of $\tnabla$ can
be extended to the whole universal exterior algebra allowing the definition of
the curvature $R(\tnabla)$ as
\begin{equation}
R(\tnabla)= -\tnabla^2~~~.
\end{equation}

We will consider gravity on $ \overline{\cal M} = {\cal M}\times Z_2$ where
$\cal M$ is the Euclidean space-time. On $\overline{\cal M}$ the algebra ${\cal
A}$ is taken as the product of two copies of $C^{\infty}$ functions on ${\cal
M}$
\begin{equation}
{\cal A} = C^{\infty}({\cal M})\oplus C^{\infty}({\cal M})~~~ .
\end{equation}
${\cal A}$ acts diagonally on the Hilbert space $\cal H$, whose
Dirac operator is
\begin{equation}
{D} = \left(\begin{array}{cc}
	     \tnabla\!\!\!\!/   &     \gamma_5 \phi    \\
	     \gamma_5 \phi   &     \tnabla\!\!\!\!/
      \end{array}\right)~~~.
\label{diracgrav}
\end{equation}

Let us introduce in $\Omega^1_D({\cal A})$ an orthonormal basis $e^N$ such that
\begin{equation}
<e^N,e^M> = \delta^{NM}~~~,
\label{enm}
\end{equation}
\begin{equation}
e^a = \left(\begin{array}{cc}
	     \gamma^a    &     0    \\
		0        &  \gamma^a
      \end{array}\right)~~~,
\label{ea}
\end{equation}
\begin{equation}
e^5 = \left(\begin{array}{cc}
	       0         &  \gamma_5 \\
	    -\gamma_5    &     0
      \end{array}\right)~~~,
\label{e5}
\end{equation}
where $N,M=1,\dots,5$; $\gamma^a = \gamma^\mu e^a_\mu$ and $e^a_\mu e^a_\nu =
g^E_{\mu\nu}$, with $a,\mu = 1,\dots,4$ \footnote{Remind that the Euclidean
Dirac matrices $\gamma_{\mu}$ and $\gamma_5$ obey to the following algebra:
$\gamma_{\mu}^{\dag} = -\gamma_{\mu}$, $\left\{\gamma_{\mu},\gamma_{\nu}
\right\}=-2g^E_{\mu \nu}$, $\gamma_{5}^{\dag}=\gamma_5$.}. The covariant
derivative operator $\tnabla$ is characterized by the matrix of 1-forms
$\omega^N_M$ defined by
\begin{equation}
\tnabla e^N = -\omega^N_M \otimes e^M~~~.
\label{tnabla}
\end{equation}
Analogously the curvature is characterized by the 2-forms $R^N_M$ given by
\begin{equation}
R(\tnabla)e^N = R^N_M \otimes e^M = (d\omega^N_M +
\omega^N_P\omega^P_M) \otimes e^M ~~~.
\end{equation}
The forms $\omega^N_M$ can be written as
\begin{equation}
\omega^N_M = \left(\begin{array}{cc}
	     \gamma^\mu~ \omega^N_{1\mu M}   &     \gamma_5\phi~ l^N_M    \\
				  & \\
	     \gamma_5\phi ~(l^T)^N_M  &    \gamma^\mu~ \omega^N_{2\mu M}
      \end{array}\right) ~~~.
\end{equation}

By imposing the conditions \eqn{lc} and \eqn{torsion} one can see that
$\omega^N_M$ can be written in terms of the Levi-Civita connection
$\omega^a_{\mu b}$ on ${\cal M}$, of the metric field $\phi$, and of the
auxiliary fields $l^N_M$ \cite{Zurgrav}. The Euclidean Einstein-Hilbert action
is
\begin{equation}
I^E_{E-H} = (32 \pi G)^{-1}\int_{\cal M} Tr[R^N_Me^M(e_N)^*] -
2\Lambda\int_{\cal M} \sqrt{-g}d^4x~~~.
\end{equation}
Eliminating the auxiliary fields and continuing to the Lorentzian
signature one obtains
\begin{equation}
I_{E-H} = \int_{\cal M} \sqrt{-g}~\left(-{1 \over 16 \pi G}
{\cal R}-2\Lambda +{ 1 \over 2} \partial_\mu\sigma\partial^\mu\sigma
\right)~d^4x~~~,
\label{actiongrav}
\end{equation}
where ${\cal R}$ stands for the scalar curvature of ${\cal M}$, and $\sigma$ is
a real scalar field such that $\phi = \mu \exp(-k\sigma)$ ($\mu$ is an
arbitrary mass and $k \equiv \sqrt{4 \pi G}= \sqrt{4 \pi/M_{Pl}^2}$).

\subsection{Gauge theory}

In this subsection we use the method just developed to add to the gravity
action the Yang-Mills action for a nonabelian gauge group
\cite{ConnesLott,Zurich}. For simplicity let us consider the case
$SU_L(n)\otimes SU_R(m)\otimes U_B(1)$ with $L$ and $R$ standing for lefthanded
and righthanded chirality, respectively. In this case the space will have the
same structure as in last section, $\overline{\cal M} \equiv {\cal M}\times Z_2
=  {\cal M}_{L} \cup {\cal M}_{R}$, with ${\cal M}$ being a smooth spin
manifold.\\We will consider the  $K$-cycle for which ${\cal H}$ is the Hilbert
space made of fermions $\psi_{L}^{b}$ and $\psi_{R}^{b}$ belonging to $n$-plet
and $m$-plet of the gauge groups $U_L(n)$ and $U_R(m)$, respectively. Hence a
vector $\psi$ of ${\cal H}$ will have the form
\begin{equation}
\Psi = \left( \begin{array}{c} \psi_{L}^{b} \\ \psi_{R}^{b}
\end{array} \right)~~~,
\label{spinor}
\end{equation}
$b=1,..,n_{g}$ standing for the family number, with $n_{g}$ the number of
generations. Thus the Hilbert space results to be
${\cal H}=L^{2}(S_{L})
\oplus L^{2}(S_{R})$ with $S_{L(R)}$ denoting a bundle on ${\cal M}_{L(R)}$
of multiplets spinors under the gauge group $U_L(n)(U_R(m))$.
The $*$-algebra, ${\cal A}$, of bounded operators acting on
${\cal H}$, is taken as
\be
{\cal A}=C^{\infty}({\cal M}) \otimes \left(U_L(n)\oplus U_R(m)\right)~~~.
\label{algebra}
\ee
An arbitrary element of ${\cal A}$, $a$, results to
be a $(n + m)\times (n+m)$ matrix
\begin{equation}
a = \left( \begin{array}{cc} a^{L} & 0 \\
0  & a^{R} \end{array} \right)~~~,
\label{aelem}
\end{equation}
where $a_L$ and $a_R$ are two unitary $n \times n$ and $m \times m$ matrices
of functions, respectively. The Euclidean Dirac operator, $D$, is given by
\begin{equation}
D \equiv \left( \begin{array}{cc} \partial\!\!\!/ \otimes \I_{n} \otimes
\I_{n_{g}}
& \gamma_{5} \phi \otimes M \otimes K\\
\gamma_{5} \phi \otimes M^{\dag} \otimes K^{\dag}&
\partial\!\!\!/ \otimes \I_{m} \otimes \I_{n_{g}}
\end{array} \right)~~~,
\label{dirac}
\end{equation}
with $M$ an $n\times m$ mass matrix, and
$K$ a matrix responsible for a family number mixing.

In this formalism the gauge potentials $A^{L(R)}_{\mu}$,
defined on the manifold ${\cal M}_{L(R)}$, are contained in the $1$-form
\begin{equation}
\rho \equiv \sum_{j} a_{j} [D,b_{j}] =
\left( \begin{array}{cc} {A\!\!\!/}^{L}  & \gamma_{5} \Phi
\otimes K \\
\gamma_{5} \Phi^{\dag} \otimes K^{\dag} & {A\!\!\!/}^{R}
\end{array} \right)~~~,
\label{rho}
\end{equation}
where in (\ref{rho}) we have used the following definitions
\begin{eqnarray}
{A\!\!\!/}^{L(R)} &=& \gamma^{\mu} A^{L(R)}_{\mu}
\equiv \gamma^{\mu} \sum_{j} a^{L(R)}_{j} \partial_{\mu}
b^{L(R)}_{j}~~~,\label{gauge}\\
\Phi &=& \phi \sum_{j} a^{L}_{j} (M b^{R}_{j} -
b^{L}_{j} M)~~~.\label{Higgs}
\end{eqnarray}
By imposing the condition
\begin{equation}
\sum_{j} a^{L(R)}_j b^{L(R)}_j =1~~~,
\label{condition}
\end{equation}
it follows that $A^{L(R)}_{\mu}$ defined in (\ref{gauge})
are antihermitian $n\times n \ (m\times m)$ matrices, which
are connected to the generators of $U_L(n)$ ($U_R(m)$). Thus in order to
obtain the gauge group $SU_L(n)\otimes SU_R(m)\otimes U_B(1)$
one has to require a graded trace
condition on $\rho$, namely $Tr (A^L_\mu)=Tr (A^R_\mu)$. We can express
$A_{\mu}^{L(R)}$ in terms of the generators\footnote{For the generators we
assume the normalization
conditions $Tr({\cal T}_{p}^{L} {\cal T}_{p'}^{L}) = 2 \delta_{p p'}$ and
$Tr({\cal T}_{q}^{R} {\cal T}_{q'}^{R}) = 2 \delta_{q q'}$.}
of $SU_L(n)$ ($SU_R(m)$)
${\cal T}_{p}^{L}$ (${\cal T}_{q}^{R}$) ($p=1,..,n^2-1$; $q=1,..,m^2-1$)
\begin{equation}
A^{L}_{\mu} \equiv - {i \over \sqrt{2}} g_{L}~ {\cal T}_{p}^{L}~
W_{ \mu}^{pL} - {i \over \sqrt{2}} g_{B} \sqrt{ \frac{m}{n(n+m)}}
{}~\I_n ~B_{\mu} ~~~,
\label{gaugepotL}
\end{equation}
\begin{equation}
A^{R}_{\mu} \equiv - { i \over \sqrt{2}} g_{R} {\cal T}_{q}^{R} W_{\mu}^{qR} -
{i \over \sqrt{2}} g_{B} \sqrt{ \frac{n}{m(n+m)}} ~\I_m~ B_{\mu} ~~,
\label{gaugepotR}
\end{equation}
where $W^{pL}_{\mu}$ ($W^{qR}_{\mu}$) are the corresponding hermitian gauge
potentials and $B_{\mu}$ is the $U_B(1)$ gauge field.
The Euclidean Yang-Mills action, as well-known, is given by
the following expression
\begin{equation}
I^{E}_{Y-M} = {1 \over 8 {\cal N}} \int d^4x \sqrt{g^E}~Tr(\theta^2)~~~,
\label{action1}
\end{equation}
where the $2$-form $\theta \equiv d\rho+\rho^2$, $d\rho=
\sum_{j} [D, a_{j}] [D, b_{j}]$, and ${\cal N}$ is a normalization constant to
be determined. From the above considerations we
get the following expressions for the matrix elements of $\theta$
\begin{eqnarray}
\theta_{LL}&=& {1 \over 2} \gamma^{\mu \nu}
G^{L}_{\mu \nu}  + \tnabla^{\mu}_L
A_{\mu}^L - \sum_{j} a^L_j {\cstok{\ }} b^L_j + \left(
\Phi \Phi^{\dag}
- \phi^2 \sum_{j} a^L_j MM^{\dag} b^L_j  \right) K K^{\dag}
{}~,\nonumber\\
\theta_{RR}&=& {1 \over 2} \gamma^{\mu \nu}
G^{R}_{\mu \nu} +  \tnabla^{\mu}_R
A_{\mu}^R - \sum_{j} a^R_j {\cstok{\ }} b^R_j + \left(
\Phi^{\dag} \Phi- \phi^2
\sum_{j} a^R_j  M^{\dag}M b^R_j  \right) K^{\dag} K
{}~,\nonumber\\
\theta_{LR} &=& \gamma_{5} \left[ -{\partial}\!\!\!/ \Phi
-k \Phi {\partial}\!\!\!/\sigma  + \Phi
{A\!\!\!/}^{R}- {A\!\!\!/}^{L} \Phi \right]K~~~,\nonumber\\
\theta_{RL} &=& \gamma_5 \left[  -{\partial}\!\!\!/ \Phi^{\dag}
- k \Phi^{\dag} {\partial}\!\!\!/\sigma
-{A\!\!\!/}^{R} \Phi^{\dag} + \Phi^{\dag} {A\!\!\!/}^{L}\right]
K^{\dag}~~~.
\label{theta1}
\end{eqnarray}
Note that, for sake of simplicity, in Eqs. (\ref{theta1}) we have shifted
the field $\Phi$ with respect to the definition (\ref{Higgs}),
and we have denoted with $\gamma^{\mu \nu} \equiv
[\gamma^{\mu},\gamma^{\nu}]/2$. Moreover, given the covariant derivative
(\ref{tnabla}), we define the gauge covariant derivative as
$\tnabla_{L(R)}^{\mu} \equiv \tnabla^{\mu}+A_{L(R)}^{\mu}$ and the Yang-Mills
tensor $G^{L(R)}_{\mu \nu} \equiv \tnabla^{L(R)}_{\mu} A^{L(R)}_{\nu}-
\tnabla^{L(R)}_{\nu} A^{L(R)}_{\mu}$. It is now convenient to introduce the
following auxiliary fields
\begin{eqnarray}
X_{L(R)} & \equiv & - \tnabla^{\mu}_{L(R)} A_{\mu}^{L(R)} +
\sum_{j} a^{L(R)}_j {\cstok{\ }} b^{L(R)}_j~~~,
\label{xlr}\\
Y_{L} & \equiv & \sum_{j} a^{L}_j  M M^{\dag} b^{L}_j~~~,
\label{yl}\\
Y_{R} & \equiv & \sum_{j} a^{R}_j  M^{\dag}M b^{R}_j~~~.
\label{yr}
\end{eqnarray}
Thus, substituting (\ref{xlr})-(\ref{yr}) in (\ref{theta1}) we get
\begin{eqnarray}
\theta_{LL}&=& {1 \over 2} \gamma^{\mu \nu}
G^{L}_{\mu \nu}  - X_L + \left(
\Phi \Phi^{\dag} - \phi^2 Y_L \right) K K^{\dag}
{}~,\nonumber\\
\theta_{RR}&=& {1 \over 2} \gamma^{\mu \nu}
G^{R}_{\mu \nu} - X_R  + \left(
\Phi^{\dag} \Phi - \phi^2 Y_R \right) K^{\dag} K
{}~,\nonumber\\
\theta_{LR} &=& \gamma_{5} \left[ -{\partial}\!\!\!/ \Phi
-k \Phi {\partial}\!\!\!/\sigma  + \Phi
{A\!\!\!/}^{R}- {A\!\!\!/}^{L} \Phi \right]K~~~,\nonumber\\
\theta_{RL} &=& \gamma_5 \left[  -{\partial}\!\!\!/ \Phi^{\dag}
- k \Phi^{\dag} {\partial}\!\!\!/\sigma
- {A\!\!\!/}^{R} \Phi^{\dag} + \Phi^{\dag} {A\!\!\!/}^{L})\right]
K^{\dag}~~~.
\label{theta2}
\end{eqnarray}
The Euclidean bosonic lagrangian density corresponding to the Yang-Mills
contribution (\ref{action1}) is then given by
\begin{eqnarray}
{\cal L}^{E}_B & = & {\sqrt{g^{E}} \over {\cal N}} \left\{
{ 1 \over 2} Tr\left[ K K^{\dag}\left( \Phi \Phi^{\dag} - \phi^2 Y_L
\right)   - X_L \right]^2   + { 1 \over 2} Tr\left[ K^{\dag} K\left(
\Phi^{\dag} \Phi- \phi^2 Y_R \right)   - X_R \right]^2 \right.
\nonumber\\
& - &  Tr\left(K K^{\dag}\right)~Tr\left[
\left(\partial_{\mu} \Phi - \Phi A^{R}_{\mu} + A^{L}_{\mu}\Phi
+ k \Phi \partial_{\mu} \sigma\right) \right.
\nonumber\\
& \times & \left. \left.
\left(\partial^{\mu} \Phi^{\dag} + A^{R\mu}
\Phi^{\dag}   -\Phi^{\dag} A^{L\mu}
+ k \Phi^{\dag} \partial^{\mu} \sigma\right) \right]
- { 1 \over 4} n_{g}~ Tr\left[ G_{L}^{\mu \nu} G^{L}_{\mu \nu} + G_{R}^{\mu
\nu} G^{R}_{\mu \nu} \right] \right\}.
 \label{action2}
\end{eqnarray}
The continuation to the {\it Lorentzian} signature is easily obtained by
substituting $ {\cal L }^{E}_B \Rightarrow - {\cal L}_B$ and
$x_{4} \Rightarrow it$.
Replacing in (\ref{action2}) the fields $A_{\mu}^{L(R)}$ with
the physical gauge potentials $W_{\mu}^{pL}$, $W_{\mu}^{qR}$ and $B_{\mu}$
we get
\begin{eqnarray}
{\cal L}_B  & = &{\sqrt{-g} \over {\cal N}} \left\{-
{ 1 \over 2} Tr\left[ K K^{\dag}\left( \Phi \Phi^{\dag} - \phi^2 Y_L
\right)   - X_L \right]^2   - { 1 \over 2} Tr\left[ K^{\dag} K\left(
\Phi^{\dag} \Phi- \phi^2 Y_R \right)   - X_R \right]^2 \right.
\nonumber\\
&+& Tr\left(K^{\dag} K\right)~Tr\left[\left({\cal D}^{\mu} \Phi
\right)^{\dag} \left({\cal D}_{\mu} \Phi \right) + k \partial _{\mu}
( \Phi^{\dag} \Phi) ~\partial^{\mu} \sigma +
k^2 \Phi^{\dag} \Phi \partial _{\mu} \sigma \partial^{\mu} \sigma
\right]
\nonumber\\
&-& \left. { 1 \over 4} n_{g}~ \left( g_{L}^2 W_{pL}^{\mu \nu} W^{pL}_{\mu \nu}
+ g_{R}^2 W_{qR}^{\mu\nu} W^{qR}_{\mu\nu}+ g_{B}^2 B_{\mu \nu} B^{\mu \nu}
\right)\right\}~~~.
\label{action4}
\end{eqnarray}
where  $W^{qL(R)}_{\mu \nu}$ and $B_{\mu \nu}$ are the Yang-Mills tensors for
left (right) and $U_B(1)$ gauge fields, respectively, and
\begin{equation}
{\cal D}_{\mu} \Phi \equiv
\partial_{\mu} \Phi + i {g_{R}\over \sqrt{2}} \Phi {\cal T}_{q}^{R}
W^{qR}_{\mu} -i {g_{L} \over \sqrt{2}} {\cal T}_{p}^{L} \Phi W^{pL}_{\mu}
+ i { g_B \over \sqrt{2}} {n-m \over \sqrt{nm (n+m)}} B_{\mu} \Phi ~~~.
\label{covdev}
\end{equation}
Note that, in order to have in Eq.(\ref{action4}) the usual kinetic terms for
the gauge degrees of freedom it is necessary to put $g_{L}=g_{R}=g_{B}$. To
obtain the final expression for the bosonic action we need to eliminate from
(\ref{action4}) the contribution coming from the auxiliary fields $Y_{L(R)}$
and $X_{L(R)}$, and this is performed by varying the lagrangian with respect to
them and substituting the resulting constraints in (\ref{action4}). It is
immediately clear from this procedure that the presence, in the final
expression of ${\cal L}_B$, of potential terms for the $\Phi$ field is strongly
depending on the form of the $M$ matrix. To be more precise, it is connected to
the possibility, for the particular choice made for $M$, of not having
independent auxiliary fields.

\subsection{A simple abelian model}

As clear from the previous sections, the scalar fields content of a
noncommutative Einstein -Yang-Mills theory is essentially independent of the
particular gauge group chosen. Any models will have at least a real $\sigma$
field coming from gravity and one or more $\Phi$ Higgs fields. This simple
consideration suggests to consider the simplest gauge theory, that is $n=m=1$,
in order to study the resulting cosmological implications. The point we wish to
make is that the corresponding results will reliably sketch the inflationary
scenario of any more complicate but more realistic GUT model.

According to Eq. (\ref{dirac}), the new Dirac operator in the Euclidean space
results to be
\begin{equation}
D \equiv \left( \begin{array}{cc} \partial\!\!\!/ \otimes
\I_{n_{g}}
& \gamma_{5}\phi  \otimes K\\
\gamma_{5}\phi  \otimes K^{\dag}&  \partial\!\!\!/ \otimes \I_{n_{g}}
\end{array} \right)~~~.
\label{diracab}
\end{equation}
Consequently, following the previous analysis we get the expression
for the matrix elements of $\theta$, which read
\begin{eqnarray}
\theta_{LL}&=& -{i \over 4} g_B~\gamma^{\mu \nu}
B_{\mu \nu}  - X_L + \left(
|\Phi|^2  - \mu^2 \exp\left( -2 k \sigma\right)
\right) K K^{\dag}
{}~,\nonumber\\
\theta_{RR}&=& - {i \over 4}g_B~ \gamma^{\mu \nu}
B_{\mu \nu} - X_R  + \left(
|\Phi|^2 - \mu^2 \exp\left( -2 k \sigma\right)
\right) K^{\dag} K
{}~,\nonumber\\
\theta_{LR} &=& - \gamma_{5} \left({\partial}\!\!\!/ \Phi +
k \Phi {\partial}\!\!\!/\sigma\right)~~,\nonumber\\
\theta_{RL} &=& - \gamma_5 \left({\partial}\!\!\!/ \Phi^{*} +
k \Phi^* {\partial}\!\!\!/\sigma\right)~~,
\label{thetaab}
\end{eqnarray}
where $\Phi$ is an usual complex field. Thus we get
the bosonic lagrangian in Lorentzian space-time which is
\begin{eqnarray}
{\cal L}_B  & =& {\sqrt{-g}\over {\cal N}} \left\{
- { 1 \over 4} n_{g} g_B^2 ~ B_{\mu \nu} B^{\mu \nu}
- \left[ Tr(K^{\dag} K)^2 - { (Tr(K^{\dag} K))^2 \over n_{g}}\right]~
\left[ |\Phi|^2 - \mu^2 \exp\left(-2 k \sigma\right)\right]^2   \right.
\nonumber\\
&+& \left. Tr\left(K^{\dag} K\right)~\left[{\partial}^{\mu} \Phi^{*}
{\partial}_{\mu} \Phi + k \partial _{\mu} |\Phi|^2 \partial^{\mu} \sigma +
k^2 |\Phi|^2 \partial_{\mu} \sigma \partial^{\mu} \sigma \right]
\right\}~~~.
\label{lagrab}
\end{eqnarray}
{}From (\ref{lagrab}) it is clear that only for $K^{\dag} K \neq \beta
{}~\I_{n_g}$ the potential term for the Higgs field $\Phi$ is preserved. With
${\cal N} = n_{g} {g_B}^2$ we obtain the usual lagrangian. Furthermore,
redefining $\Phi \equiv \varphi~ g_B \left[n_{g} /  Tr(K^{\dag}K)
\right]^{1/2}$, and finally $\mu \equiv m~ g_B \left[n_{g} / Tr(K^{\dag}K)
\right]^{1/2}$ we get
\begin{eqnarray}
{\cal L}_{B}  & = &\sqrt{-g} \left\{
- { 1 \over 4} B_{\mu \nu} B^{\mu \nu}
- {\lambda \over 4!} \left[ |\varphi|^2 -
m^2 \exp\left(- 2 k \sigma\right)\right]^2 - V_0   \right.
\nonumber\\
&+& \left. \left[{\partial}^{\mu} \varphi
{\partial}_{\mu} \varphi^* +  k
\partial _{\mu} |\varphi|^2 ~\partial^{\mu} \sigma +
\left({1\over2} + k^2 |\varphi|^2 \right)
\partial_{\mu} \sigma \partial^{\mu} \sigma \right]\right\}~~~.
\label{lagrab1}
\end{eqnarray}
In (\ref{lagrab1}) the kinetic term of the $\sigma$-field and the
constant cosmological term appearing in (\ref{actiongrav}),  $V_0$,
are also included, and the definition
\begin{equation}
{\lambda} \equiv 6~ g_B^2
\left[ {n_{g} Tr(K^{\dag} K)^2 \over (Tr(K^{\dag} K))^2} -1 \right] \geq 0
\label{lambda}
\end{equation}
is used. From Eq. (\ref{lagrab1}) one easily
gets the tree-level potential for the $\varphi$ and $\sigma$ fields
$V(\sigma,\varphi)$
\begin{equation}
V(\sigma,\varphi) = V_0 + {\lambda \over 4!} \left[ |\varphi|^2 -
m^2 \exp\left(- 2 k \sigma\right)\right]^2~~~.
\label{treepot}
\end{equation}
This potential has its minimum for $|\varphi|^2
= m^2 \exp\left(- 2 k \sigma\right)$,
which is a condition which does not fix both $|\varphi|$ and $\sigma$
uniquely.
The situation changes if one assumes $\sigma$ to be still a
classical field, and consider
the one-loop corrections to $V(\sigma,\varphi)$ coming from the quantum
fluctuations of $\varphi$ \cite{Coleman,Ringwald}.
This assumption is consistent with the physical
scenario we want to describe, since the inflationary epoch
is assumed to be described by Einstein gravity (classical theory).
In our case this means that the classical Einstein theory can be applied to
the {\it doubled} manifold, so $\sigma$, which is the metric component in the
{\it discrete} dimension, has to be a pure classical field with no quantum
fluctuations. On the contrary, $\varphi$ is an usual scalar field and thus
for it we will consider quantum fluctuations.

\section{Inflation from scalar dynamics}

To study the inflation induced in the early universe by the dynamics
of $\sigma$ and $\varphi$, one needs to solve the equations of motion
for these fields starting from reasonable initial conditions. In this
section we will discuss the evolution of this coupled system showing
how this evolution corresponds, for a natural choice of initial
conditions, to an inflationary phase.

\subsection{The effective potential}

The Higgs field $\varphi$, contrary to $\sigma$ which we will
treat as a pure
classical field, can be expanded around its background configuration
as
\begin{equation}
\varphi = {\varphi_c \over \sqrt{2}}
+ { \delta\varphi_1 + i ~\delta\varphi_2 \over \sqrt{2}}~~~,
\label{expansion}
\end{equation}
where the background configuration $\varphi_c/\sqrt{2} \equiv \langle\varphi
\rangle$ can always be chosen real using the $U(1)$ gauge symmetry of
the theory. With $\delta\varphi_i$ we denote the
independent quantum fluctuations, which satisfy $\langle \delta\varphi_i
\rangle =0$. By substituting (\ref{expansion}) in
(\ref{lagrab1}), we get the following
equations of motion for $\varphi_c$, $\sigma$, and $\delta\varphi_i$
\begin{eqnarray}
\cstok{\ }\varphi_c + k \varphi_c \cstok{\ } \sigma - k^2 \varphi_c
\partial_{\mu} \sigma \partial^{\mu} \sigma + { \lambda \over 24}
\left[ \varphi_c^2 -2 m^2 \exp\left(- 2 k \sigma\right) +
\langle 3 ~\delta \varphi_1 ^2 + \delta \varphi^{2}_2  \rangle \right]
\varphi_c =0~~~,\nonumber\\
\label{eq1}
\end{eqnarray}
\begin{eqnarray}
\left[ 1 + k^2 \left(\varphi_c^2 + \langle
\delta \varphi_1 ^2 + \delta \varphi^{2}_2  \rangle
\right)\right] \cstok{\ } \sigma
+ 2 k^2 \varphi_c \partial_{\mu}
\varphi_c \partial^{\mu} \sigma + k  \partial_{\mu}
\varphi_c \partial^{\mu} \varphi_c + k \varphi_c \cstok{\ } \varphi_c
\nonumber\\
+ {\lambda \over 12} k~ m^2   \exp\left(-2k \sigma \right)
\left[ \varphi_c^2 -2 m^2 \exp\left(- 2 k \sigma\right) +
\langle \delta \varphi_1^2 + \delta \varphi_2^2 \rangle\right]=0~~~,
\label{eq2}
\end{eqnarray}
\begin{eqnarray}
\cstok{\ } \delta\varphi_1 +
{\lambda \over 24}
\left[ 3 \varphi_c^2 -2 m^2 \exp\left(- 2 k \sigma\right)\right]
\delta\varphi_1 + {\lambda \over 24} \left[
\delta \varphi_1^3 + \delta \varphi_1
\delta \varphi_2^2 \right]
\nonumber\\
+ {\lambda \over 24} \varphi_c \left[ 3 \left(\delta \varphi_1^2
- \langle \delta \varphi_1^2 \rangle \right) +
\left(\delta \varphi_2^2- \langle \delta \varphi_2^2 \rangle \right)
\right]=0~~~,
\label{eq3}
\end{eqnarray}
\begin{eqnarray}
\cstok{\ } \delta\varphi_2 +
{\lambda \over 24}
\left[ \varphi_c^2 -2 m^2 \exp\left(- 2 k \sigma\right)\right]
\delta\varphi_2
+ {\lambda \over 24} \left[\delta \varphi_2^3 + \delta \varphi_1^2
\delta \varphi_2 + 2 \varphi_c \delta \varphi_1 \delta \varphi_2
\right]=0~.
\label{eq4}
\end{eqnarray}
The quantities $\langle \delta \varphi_i^2 \rangle$ in Eqs. (\ref{eq1}) and
(\ref{eq2}) account for the one-loop corrections to the effective
potential.
Note that in the quantum fluctuation equations of
motion (\ref{eq3}) and (\ref{eq4})
the equation of motion for $\varphi_c$ has been used, and
according to the adiabatic approximation \cite{Ringwald},
we have neglected all terms proportional to derivatives of the
background fields.\\
In order to linearize the last two equations with respect
to $\delta\varphi_i$, one can estimate the self-interaction terms of
these fields in the {\it mean-field} approximation. According to
this ansatz, the quadratic and cubic terms of $\delta\varphi_i$
result to be
$\delta \varphi_i^3 = 3\langle \delta \varphi_i^2 \rangle
\delta \varphi_i$, $\delta \varphi_i^2 = \langle \delta \varphi_i^2
\rangle$ and
$\delta \varphi_1 \delta \varphi_2 =
\langle \delta \varphi_1 \delta \varphi_2 \rangle = 0$ respectively.
In particular the vanishing of the $\delta \varphi_1
\delta \varphi_2$ correlator
can be easily understood from the explicit expression of the interaction
term of the fluctuations; from (\ref{treepot})
\begin{equation}
V_{\delta \varphi_1 \delta \varphi_2} = \frac{\lambda}{96}
\left[ 4 \varphi_c
( \delta \varphi_1^3 + \delta \varphi_1 \delta \varphi_2^2 )
+ 2 \delta \varphi_1^2 \delta \varphi_2^2 +
\delta \varphi_1^4 + \delta \varphi_2^4 \right]
\label{eqvarpot}
\end{equation}
one get the conclusion that, at all orders in perturbation theory, there
are no contributions to graphs with an odd number of $\delta \varphi_2$
external lines so, in particular, $\langle \delta \varphi_1
\delta \varphi_2 \rangle=0$. We also notice that all expectation values of
cubic terms, like $\langle \delta \varphi_1^3 \rangle$, do not contribute to
(\ref{eq3}) and (\ref{eq4}). By taking the expectation value of these
equations, in fact, and using the condition $\langle \delta \varphi_i
\rangle=0$,
it follows that the sum of all these contributions
identically vanishes.\\
The linearized expressions for Eqs. (\ref{eq3}) and (\ref{eq4}) take
then the form
\begin{eqnarray}
\cstok{\ } \delta\varphi_1 +
{\lambda \over 24}
\left[ 3 \varphi_c^2 -2 m^2 \exp\left(- 2 k \sigma\right)
+ \langle 3~\delta \varphi_1^2 + \delta \varphi_2^2\rangle\right]
\delta\varphi_1 =0~~~,
\label{eq3a}
\end{eqnarray}
\begin{eqnarray}
\cstok{\ } \delta\varphi_2 +
{\lambda \over 24}
\left[ \varphi_c^2 -2 m^2 \exp\left(- 2 k \sigma\right)
+ \langle \delta \varphi_1^2 + 3~\delta \varphi_2^2\rangle\right]
\delta\varphi_2
=0~~~.
\label{eq4a}
\end{eqnarray}
By following \cite{Ringwald}, at zero-th order in the adiabatic expansion,
we get the {\it gap} equations for $\langle \delta \varphi_i^2 \rangle$
\begin{eqnarray}
\langle \delta \varphi_1^2 \rangle = { 1 \over 8 \pi^2 }
\left\{ Q^2 + \left[
 {\lambda \over 48}
\left( 3 \varphi_c^2 -2 m^2 \exp\left(- 2 k \sigma\right)
+ 3 \langle \delta \varphi_1^2 \rangle + \langle \delta \varphi_2^2 \rangle
\right) \right] \right.\nonumber\\
\times \left.
\left[\log \left| { \lambda
\left[ 3 \varphi_c^2 -2 m^2 \exp\left(- 2 k \sigma\right)
+ 3 \langle \delta \varphi_1^2 \rangle + \langle \delta
\varphi_2^2 \rangle \right] \over 96 Q^2}\right|
+ 1 \right] \right\} ~~~,
\label{gapdphi12}
\end{eqnarray}
\begin{eqnarray}
\langle \delta \varphi_2^2 \rangle = { 1 \over 8 \pi^2 }
\left\{ Q^2 + \left[
{\lambda \over 48}
\left( \varphi_c^2 -2 m^2 \exp\left(- 2 k \sigma\right)
+  \langle \delta \varphi_1^2 \rangle + 3
\langle \delta \varphi_2^2 \rangle
\right) \right] \right.\nonumber\\
\times \left.
\left[\log \left| { \lambda
\left[ \varphi_c^2 -2 m^2 \exp\left(- 2 k \sigma\right)
+  \langle \delta \varphi_1^2 \rangle + 3
\langle \delta \varphi_2^2 \rangle
 \right] \over 96 Q^2}\right|
+ 1 \right] \right\} ~~~.
\label{gapdphi22}
\end{eqnarray}
Note that for simplicity, we have neglected the contribution of the coupling
of $\varphi$ with the scalar curvature $R$, because it rapidly becomes
negligible. Thus the Eqs. (\ref{gapdphi12}) and (\ref{gapdphi22})
are just the flat space corrections {\it \`a la} Coleman-Weinberg \cite{
Coleman}.

The parameter $Q$ is the ultraviolet cut-off which, representing
the energy upper bound for the validity of our model, will
be assumed in the following of the order of the Planck mass.
These two expressions can be solved by iteration. In the following we
will only consider the zero-th order approximation, which already gives
results correct at order $\lambda$
\begin{eqnarray}
\langle \delta \varphi_1^2 \rangle = { 1 \over 8 \pi^2 }
\left\{ Q^2 + \left[
{\lambda \over 48}
\left( 3 \varphi_c^2 -2 m^2 \exp\left(- 2 k \sigma\right)
+ { Q^2 \over 2 \pi^2 }\right) \right] \right.\nonumber\\
\times \left.
\left[\log \left| { \lambda
\left[ 3 \varphi_c^2 -2 m^2 \exp\left(- 2 k \sigma\right)
+ Q^2 / (2 \pi^2 )\right]  \over 96 Q^2}\right|
+ 1 \right] \right\} ~~~,
\label{dphi12}
\end{eqnarray}
\begin{eqnarray}
\langle \delta \varphi_2^2 \rangle = { 1 \over 8 \pi^2 }
\left\{ Q^2 + \left[
{\lambda \over 48}
\left( \varphi_c^2 -2 m^2 \exp\left(- 2 k \sigma\right)
+ { Q^2 \over 2 \pi^2 }\right) \right] \right.\nonumber\\
\times \left.
\left[\log \left| { \lambda
\left[ \varphi_c^2 -2 m^2 \exp\left(- 2 k \sigma\right)
+ Q^2 /(2 \pi^2 )\right] \over 96 Q^2}\right|
+ 1 \right] \right\} ~~~.
\label{dphi22}
\end{eqnarray}
{}From Eqs. (\ref{eq3a}) and (\ref{eq4a}) one reads the square of
the effective masses for the quantum fluctuations
\begin{eqnarray}
\Omega_1^2 & = &
{\lambda \over 24}
\left[3 \varphi_c^2 -2 m^2 \exp\left(- 2 k \sigma\right)
+ { Q^2 \over 2 \pi^2 } \right]~~~,\label{eqomega12}\\
\Omega_2^2 & = &
{\lambda \over 24}
\left[ \varphi_c^2 -2 m^2 \exp\left(- 2 k \sigma\right)
+ { Q^2 \over 2 \pi^2 } \right]~~~.\label{eqomega22}
\end{eqnarray}
As it always occurs in scalar field theories, these masses get
contributions from the highest scale involved in the model,
which in this case is the Planck mass.

By virtue of Eqs. (\ref{dphi12}) and (\ref{dphi22}) one can
determine the quantum fluctuation
contributions to (\ref{eq1}) and (\ref{eq2})
\begin{eqnarray}
\cstok{\ }\varphi_c + k \varphi_c \cstok{\ } \sigma - k^2 \varphi_c
\partial_{\mu} \sigma \partial^{\mu} \sigma +
{\delta V_{eff} \over \delta \varphi_c }
=0~~~,
\label{eq1a}
\end{eqnarray}
\begin{eqnarray}
\left[ 1 + k^2 \left( \varphi_c^2 + { Q^2 \over 4 \pi^2 } \right)\right]
\cstok{\ } \sigma
+ 2 k^2 \varphi_c \partial_{\mu}
\varphi_c \partial^{\mu} \sigma + k  \partial_{\mu}
\varphi_c \partial^{\mu} \varphi_c + k \varphi_c \cstok{\ } \varphi_c
+{\delta V_{eff} \over \delta \sigma }=0,\nonumber\\
\label{eq2a}
\end{eqnarray}
where the effective potential $V_{eff}(\sigma, \varphi_c)$,
in terms of the definitions (\ref{eqomega12}) and (\ref{eqomega22}),
results to be
\begin{eqnarray}
V_{eff}(\sigma,\varphi_c) & = & V_0 + {\lambda \over 96}
\left[ \varphi_c^2 - 2 m^2 \exp\left(- 2 k \sigma\right)\right]^2 +
{\lambda Q^2 \over 96 \pi^2} \left[ \varphi_c^2 -
m^2 \exp\left(- 2 k \sigma\right)\right]
\nonumber\\
& + & { 1 \over 64 \pi^2} \left\{ \Omega^4_1 \left[
\log\left|{\Omega^2_1 \over 4 Q^2}\right| + {1 \over 2} \right] +
\Omega^4_2 \left[
\log\left|{\Omega^2_2 \over 4 Q^2}\right| + {1 \over 2} \right]\right\}
{}~~~.
\label{effpot}
\end{eqnarray}
Notice that in the factor in front of $\cstok{\ } \sigma$ in Eq. (\ref{eq2a})
we have only considered the zero-th order term in the $\lambda$ expansion;
this is to be consistent with the fact that, according to the adiabatic
expansion, we have neglected in Eqs. (\ref{eq3}) and (\ref{eq4})
the terms order $\lambda$, which are proportional to the derivative of the
$\sigma$ field.

The quadratic and logarithmic divergences in $V_{eff}$ when $Q^2 \rightarrow
\infty$ are usually removed by the renormalization procedure, namely by a
redefinition of the mass and the coupling constant. This is always possible for
a potential of the form (\ref{treepot}) if the term $m^2 \exp(-2 k \sigma)$ is
just a constant. In our case, however, the tree level expectation value of the
$\varphi$ field depends on the dynamical field $\sigma$. If one assumes that
the latter evolves quite rapidly with respect to $\varphi$, once $\sigma$ has
reached a stationary point $\sigma_m$, the dynamics of the $\varphi$ becomes
the one governed by the usual {\it mexican hat} potential
\begin{equation}
V(\sigma,\varphi) = V_0 + {\lambda \over 4!} \left[ |\varphi|^2 -
m^2 \exp\left(- 2 k \sigma_m\right)\right]^2~~~,
\label{treepot2}
\end{equation}
which can be renormalized in order to take into account the effects of quantum
fluctuations. In the general case we are dealing with, where the simultaneous
evolution of $\varphi$ and $\sigma$ is considered, our model is not
renormalizable, i.e. both divergences cannot simply absorbed in the parameters
of the lagrangian density. This fact is not surprising, since the $\sigma$
field is related to gravity in the discrete direction, and still we are far
{}from a renormalizable theory where gravitational fields are considered as
dynamical degrees of freedom. In the following we will look at (\ref{effpot})
as an effective potential and we will fix the cut-off, as already mentioned, at
the order of the Planck mass scale, which is the natural one at which the field
dynamics takes place. We will see, by numerically solving the one-loop
equations of motions, that in fact the evolution of $\sigma$ is quite fast due
to the exponential nature of its potential. However, the largest contribution
to the $e$-fold number just comes from the time interval during which $\sigma$
reaches its stationary point, so the $\sigma$ dynamics, as far as the
inflationary dynamics is concerned, is particularly relevant.

The study of the minima of the effective potential (\ref{effpot}) can be
analytically performed in a simple way if we consider only terms up to the
first order in $\lambda$: in this case the only minimum is located at the point
$\varphi_{c,m}=0$, $\sigma_m= (1/2k)~\log(8 \pi^2 m^2/Q^2)$. By requiring that
the effective cosmological constant vanishes at the absolute minimum we
therefore get $V_0= \lambda Q^4/(1536 \pi^4)$. We have numerically checked that
the shift in the true minimum due to the order $\lambda^2$ logarithmic terms is
of few percents for $\lambda \sim .1 $, which will be the largest value we will
use in the following in order to have a reliable perturbative treatment of the
scalar dynamics.

In Figure 1. the effective potential for a choice of parameters, namely
$Q=10~M_{Pl}$, $m=1~M_{Pl}$ and $\lambda=0.1$, is shown.

\subsection{The equations of motion}

Since we are interested in the cosmological implications of the scalar
dynamics, and in particular in the possibility to get a reasonable inflationary
stage, we choose a spatially flat Friedmann-Robertson-Walker (FRW) universe,
whose metric has the simple form
\be
ds^2 = dt^2 - a^2(t) ~d\vec{x}^2~~~,
\label{FRWmetric}
\ee
and assume homogeneous field configurations, namely $\sigma=\sigma(t)$,
$\varphi_c=\varphi_c(t)$.\\ Under this assumption, the equations of motion
Eqs.(\ref{eq1a})-(\ref{eq2a}), once put in normal form, read
\begin{eqnarray}
\ddot{\sigma}+3 H \dot{\sigma} +
\left(1 + {k^2 Q^2 \over 4 \pi^2} \right)^{-1}
\left[ k (\dot{\varphi}_c+ k \varphi_c
\dot{\sigma})^2
+{\delta V_{eff} \over \delta \sigma }
- k \varphi_c {\delta V_{eff} \over \delta \varphi_c }\right]=0~~~,
\label{eqsigma1}
\end{eqnarray}
\begin{eqnarray}
\ddot{\varphi}_c + 3 H \dot{\varphi}_c - k \varphi_c
\left(1 + {k^2 Q^2 \over 4 \pi^2} \right)^{-1}
\left[ k (\dot{\varphi}_c+ k \varphi_c \dot{\sigma})^2
+{\delta V_{eff} \over \delta \sigma }
- k \varphi_c {\delta V_{eff} \over \delta \varphi_c }\right]
\nonumber\\
- k^2 \varphi_c \dot{\sigma}^2 + {\delta V_{eff} \over \delta \varphi_c }
=0~~~,
\label{eqvarphi1}
\end{eqnarray}
where $H$ is the Hubble parameter, defined as
\begin{eqnarray}
H \equiv \left( { \dot{a} \over a } \right)
= \sqrt{{2 \over 3}} k ~\left\{ {\dot{\varphi}_c^2 \over 2}
+ k \varphi_c \dot{\varphi}_c \dot{\sigma} + { 1 \over 2}
\left[ 1 + k^2 \left( \varphi_c^2 + { Q^2 \over 4 \pi^2} \right)\right]
\dot{\sigma}^2 + V_{eff}(\sigma,\varphi_c)
\right\}^{1/2}~~~.
\label{hubble}
\end{eqnarray}
We will fix the initial conditions for $\sigma$ and $\varphi_c$ by requiring
that the corresponding initial value for the energy density is of the order of
$M_{Pl}^4$, with vanishing kinetic terms. Moreover, we also require that the
initial distance between the two sheets, $d_5(0)$, is equal to the Planck
length, to be consistent with the idea that our model is able to describe the
evolution of the universe just after the quantum epoch for gravity. Since all
dynamics only depends on the combination $d_5^{-1}= m~\exp(-k \sigma)$, and $m$
can be reexpressed in terms of $d_5(0)$, we get $d_5^{-1} = d_5^{-1}(0)
{}~\exp[-k (\sigma-\sigma(0))]$. This allows, by shifting $\sigma \rightarrow
\sigma + \sigma(0)$, to take $\sigma(0)=0$ as initial condition without loss of
generality, since all the other terms in the equation of motion are invariant
under this transformation. Consequently, the requirement on the initial energy
density fixes, up to an inessential sign, $\varphi_c(0)$.

For the free parameters of $V_{eff}$, $\lambda$ and $Q$, we have considered the
following ranges: $\lambda \in [0.01,0.1]$, which ensures the reliability of
the perturbative approach, and $Q \gapproxeq M_{Pl}$. The latter choice is, of
course, based on the reasonable assumption that in the initial stage the only
typical mass scale is $M_{Pl}$.

The set of coupled equations (\ref{eqsigma1})-(\ref{hubble}) has been
numerically solved. We will here sketch the main features of the $\sigma-
\varphi_c$ motion: a more detailed analysis is reported in the next section,
where also the reheating epoch is considered.

The dynamics is characterized by two main behaviours: in a first phase the
scalar fields evolves towards the minimum $\sigma=\sigma_m, \varphi\sim 0$ (see
section 3.1). During this {\it rolling down}, as in the chaotic scenarios, the
scale factor rapidly grows, faster the larger is the value of the cut-off
parameter $Q$. It is interesting to note that, since the stationary point for
$\varphi$ differs from zero for terms of the order $\lambda^2$, the $U(1)$
symmetry is broken quite weakly. In particular, if the choice for $Q$ satisfies
the condition $Q < 2 \pi M_{Pl}$ it is easy to see that the $\varphi_c$ field
undergoes an {\it inverse} phase transition. At $t=0$ in fact, having chosen
$d_5(0)= M_{Pl}^{-1}$, $\Omega_i^2(\varphi_c=0)<0$, showing that we are in a
broken symmetry phase: in these conditions, in fact, the minimum at fixed
$\sigma=\sigma(0)=0$ is located at $\varphi_{c,m}^2= 2 M_{Pl}^2 - Q^2/2 \pi^2$.
When $d_5$, due to the $\sigma$ evolution, reaches the critical value $d_5^{cr}
= 2 \pi/Q> d_5(0)$ the symmetry is restored, up to terms order $\lambda^2$ due
to the logarithmic terms in $V_{eff}$, since the true minimum becomes
$\varphi_c=0$. Differently, if $Q \geq 2 \pi M_{Pl}$ all the $\varphi_c$
evolution take place in the unbroken phase, since in this case $d_5$ always
remains above the critical value.

Once the rolling down stops, the value of $\sigma$ remains practically fixed at
$\sigma_m$, while $\varphi_c$ starts oscillating around the equilibrium
configuration. This phase is extremely important for reheating since, as well
known, during these oscillations the scalar fields may transfer their energy
density to the incoherent gas of $\delta \varphi_i$ quanta and, more generally,
to other particles via decay processes.

\subsection{The Reheating}

It is well known that the {\it necessary } last stage of the inflationary
dynamics is the so-called reheating era. During this phase the energy density
stored in the scalar sector is released to the relativistic degrees of freedom
{\it via} decay processes. This epoch is at least as important as the truly
inflationary one, since it guarantees the end of the exponential-like growth of
the scale factor and the reheating of the universe, which will then continue to
expand as in a radiation dominated era. It is usually stated that the inflaton
field configuration, once it has reached its minimum, can be regarded as a
condensate of the corresponding massive quanta, which would decay in other
light particles. Such a coupling is assumed to give rise to an effective term
of the form $\Gamma \dot{\Phi}$, where $\Gamma$ is the decay rate, which is put
{\it by hand} into the inflaton equation of motion. It has been shown however
that this procedure is unsatisfactory \cite{kofman}, \cite{boyan}. A better
approximation consists in deducing the friction terms from the one-loop effects
of quantum fluctuations at first order in the adiabatic expansion, as done in
\cite{Ringwald} and \cite{boyan}. This procedure would imply the appearance of
terms of the form $\Gamma H \Phi $ in the inflaton equation of motion and,
moreover, that the mass of the $\Phi$ quanta should be considered as time
dependent. We will restrict ourselves in the following to this approach; this
will give a correct qualitative description for small times, but, as any other
perturbative computation, it would fail for large times. This is due, in
particular, to resonance phenomena which enhance the order $\hbar$ term
amplitude over the tree amplitude, so one should make use of a {\it Hartree} or
{\it mean field} technique similar to the one discussed in the previous section
\cite{boyan}. This would lead to a set of gap equations to be solved at higher
order in the adiabatic expansion. However it is outside the aim of this paper a
detailed analysis of the reheating of the universe: the final stage of the
inflationary era, of course, strongly depends on the particular model one
assumes, as well on the way the inflaton field is coupled to the relativistic
degrees of freedom. What we would like to show is simply that even the simplest
abelian model may give rise to an efficient transfer of the energy density from
the coupled system of scalar field configurations $\sigma-\varphi_c$ to
radiation.

To this end we will describe the reheating phase as due to decay processes of
the $\varphi$ field into fermionic degrees of freedom $\Psi$ (see eq.
(\ref{spinor})), due to the interaction term $Tr[\overline{\Psi} (D + \rho)
\Psi]$, where $D$ is defined by (\ref{diracab}) and $\rho$ is the abelian
version of the gauge potential matrix (\ref{rho}). These processes lead to an
energy transfer from the coherent classical configuration, whose evolution
drives the inflationary expansion of the universe, to the incoherent gas of
$\Psi$ quanta: if the latter starts to dominate the energy density of the
universe, the inflation stops, since their equation of state relating the
energy density $\rho_R$ and the pressure $p_R$ satisfies the condition $ \rho_R
+ 3 p_R = 4 \rho_R > 0 $.

The interaction term can be written in the usual Yukawa form
\begin{equation}
{\cal L}_{Yuk} = - f~ a(t)^4
\varphi~ \overline{\psi_L} \psi_R + h. c.~~~.
\label{yukawa}
\end{equation}
Using the background splitting (\ref{expansion}) we have
\begin{equation}
{\cal L}_{Yuk} = - \frac{f}{\sqrt{2}} ~ a(t)^4 \left[
\varphi_c \overline{\psi_L} \psi_R + \left( \delta \varphi_1 + i
\delta \varphi_2 \right) \overline{\psi_L} \psi_R \right]+ h. c.~~~,
\label{yukawa2}
\end{equation}
implying that the fermions acquire an effective mass $m_{\Psi} = f
\varphi_c/ \sqrt{2} $. For the decay of the $\delta \varphi_i$ quanta
into a fermion-antifermion pair to be possible the scalar particle
masses should be greater than twice the fermion mass. The
corresponding decay probabilities in the $\delta \varphi_i$ rest frame
can be simply evaluated in the Born approximation and result to be
\begin{equation}
\Gamma_i \equiv \Gamma( \delta \varphi_i \rightarrow \Psi \overline{\Psi}
) = n_f \frac{ f^2}{16 \pi} \Omega_i ~ \Theta ( \Omega_i - 2 m_{\Psi} )
{}~~~i=1,2~~~,
\label{gamma}
\end{equation}
where $n_f \sim 10^2$ is the number of independent decay channels.
In the general case of a non-abelian gauge theory, one should also
consider interaction terms of the scalar field $\Phi$ with the gauge
bosons, see (\ref{action4}). This would increase the number of
possible decay channels for $\Phi$ and the efficiency of the entropy
release to the relativistic degrees of freedom. However the analysis
for the simple abelian case will be general enough to show how the
model may also quite naturally account for the end of inflation.

The $\varphi_c$ and $\sigma$ equations of motion modify due to the
presence of the interaction terms with the fermionic field: the
expectation value of the fluctuation field squared $\langle
\delta \varphi_i^2 \rangle$ should in fact include the interaction with
the $\Psi$ degrees of freedom. Following \cite{Ringwald}, to which
for brevity we refer for the detailed calculation, one finds
\begin{equation}
\langle \delta \varphi_i^2 \rangle =
\langle \delta \varphi_i^2 \rangle_0 +{1\over {256 \pi} }\left[
\frac{\dot{\Omega}_i}
{\Omega_i} + H \right] \Gamma_i
\label{green1}
\end{equation}
where with $\langle \delta \varphi_i^2 \rangle_0$ we have denoted the
corresponding result in absence of the interaction term (\ref{yukawa2}),
which has been quoted in (\ref{dphi12}) and (\ref{dphi22}).
Using these expressions we can compute the equations of motion for
$\sigma$ and $\varphi_c$; as before, inserting (\ref{green1}) into
(\ref{eq1}) and (\ref{eq2}) and passing to normal form
\begin{eqnarray}
\ddot{\sigma}+3 H \dot{\sigma} +
\left(1 + {k^2 Q^2 \over 4 \pi^2} \right)^{-1}
\Bigr[ k (\dot{\varphi}_c+ k \varphi_c
\dot{\sigma})^2
+{\delta V_{eff} \over \delta \sigma }
- k \varphi_c {\delta V_{eff} \over \delta \varphi_c }\nonumber\\
+ \gamma_{\sigma} - k \varphi_c \gamma_{\varphi_c}\Bigr]=0~~~,
\label{eqsigma2}
\end{eqnarray}
\begin{eqnarray}
\ddot{\varphi}_c + 3 H \dot{\varphi}_c - k \varphi_c
\left(1 + {k^2 Q^2 \over 4 \pi^2} \right)^{-1}
\Bigr[ k (\dot{\varphi}_c+ k \varphi_c \dot{\sigma})^2
+{\delta V_{eff} \over \delta \sigma }
- k \varphi_c {\delta V_{eff} \over \delta \varphi_c }\nonumber\\
+ \gamma_{\sigma} - k \varphi_c \gamma_{\varphi_c} \Bigr]
- k^2 \varphi_c \dot{\sigma}^2 + {\delta V_{eff} \over \delta \varphi_c }
+ \gamma_{\varphi_c} =0~~~.
\label{eqvarphi2}
\end{eqnarray}
With $\gamma_{\varphi_c}$ and $\gamma_{\sigma}$ we have denoted the
following quantities
\begin{eqnarray}
\gamma_{\varphi_c} & = &  \frac{\lambda \varphi_c}{6144 \pi}
\left[ \left( 3 \Gamma_1 + \Gamma_2 \right) H + 3
\frac{\dot{\Omega}_1}{\Omega_1} \Gamma_1 +
\frac{\dot{\Omega}_2}{\Omega_2} \Gamma_2 \right] \label{friz1} \\
\gamma_{\sigma} & = &  \frac{\lambda k m^2 \exp(-2 k \sigma) }{3072 \pi}
\left[ \left(  \Gamma_1 + \Gamma_2 \right) H +
\frac{\dot{\Omega}_1}{\Omega_1} \Gamma_1 +
\frac{\dot{\Omega}_2}{\Omega_2} \Gamma_2 \right]
\label{friz2}
\end{eqnarray}
which represent the dissipation terms due to fermion-antifermion
pairs production.

The complete system of differential equations describing the evolution in
the inflationary phase we are interested in, includes also the equation
ruling the evolution in time of the energy density
of the relativistic fluid of $\Psi$ field quanta,
which are produced via the decay term. This equation is provided by
the covariant conservation of the energy-momentum tensor which reads
\be
d (a^3 \rho_R) = - (p_R + p_{\sigma,\varphi_c}) d(a^3) - d (
a^3 \rho_{\sigma,\varphi_c})~~~,
\label{fluideq}
\ee
where
\be
\rho_{\sigma,\varphi_c}= {\dot{\varphi}_c^2 \over 2} + k \varphi_c
\dot{\varphi}_c \dot{\sigma} + { 1 \over 2} \left[ 1 + k^2 \left(
\varphi_c^2 + { Q^2 \over 4 \pi^2} \right)\right] \dot{\sigma}^2 +
V_{eff}(\sigma,\varphi_c)~~~,
\label{rhosigphi}
\ee
\be
p_{\sigma,\varphi_c}= {\dot{\varphi}_c^2 \over 2} + k \varphi_c
\dot{\varphi}_c \dot{\sigma} + { 1 \over 2} \left[ 1 + k^2 \left(
\varphi_c^2 + { Q^2 \over 4 \pi^2} \right)\right] \dot{\sigma}^2 -
V_{eff}(\sigma,\varphi_c)~~~,
\label{psigphi}
\ee
represent the pressure and the energy density of the total system
of $\sigma-\varphi_c$ fields, respectively.

Equation (\ref{fluideq}), by using the definitions of
$\rho_{\sigma,\varphi_c}$ and $p_{\sigma,\varphi_c}$ and
the equations of motion (\ref{eqsigma2}), (\ref{eqvarphi2}),
can be cast in the form
\be
\dot{\rho}_R + 4 H \rho_R = \gamma_{\sigma} \dot{\sigma} +
\gamma_{\varphi_c} \dot{\varphi}_c~~~.
\label{fluideq2}
\ee
Finally, for the Hubble parameter $H$ we get the expression
\begin{eqnarray}
H \equiv \left( { \dot{a} \over a } \right)
= \sqrt{{2 \over 3}} k ~\left[ \rho_{\sigma,\varphi_c} +\rho_R
\right]^{1/2}~~~.
\label{hubble1}
\end{eqnarray}
The system of equations (\ref{eqsigma2}), (\ref{eqvarphi2}),
(\ref{fluideq2}) and (\ref{hubble1}) allows
to compute the $e$-fold number $N$ of the inflationary stage,
defined as $N\equiv \log\left[a(t_f)/a(0)\right]$, where
$t_f$ is the value of time at which the inflation ends. This time
can be fixed as the one at which the radiation energy $\rho_R(t_f)$
eventually equals the energy density $\rho_{\sigma,\varphi_c}(t_f)$
stored in the  background configuration $\varphi_c-\sigma$.
In terms of the Hubble parameter (\ref{hubble1}),
the expression for $N$ takes the simple form
\be
N= \int^{t_f}_{0} H(\sigma(t),\varphi(t),\rho_R(t))~ dt~~~.
\label{efold}
\ee
The reheating temperature, namely the temperature at the beginning of
radiation dominated era, can then be found by expressing the radiation
energy density $\rho_R(t_f)$ in terms of the temperature. If one
supposes in fact that a complete thermalization already occurred
at $t_f$ due to fast gauge bosons mediated interactions among
the fermions, we have
\be
\rho_R(t_f) = n_f { \pi \over 30}~T_{RH}^4~~~.
\label{trh}
\ee
In Table 1 the $e$-fold $N$ and the reheating temperature are computed
for some values of $Q$ and $\lambda$, with $f=1$, while in Figure 2 we
report the trajectories in the $\sigma-\varphi_c$ plane for the same
choices of the free parameters.

As already mentioned in the previous section the $\sigma-\varphi_c$
motion is initially characterized by a rolling down toward the
minimum. During this phase, shown in Figure 2, the radiation energy
density remains very small while the $e$-fold $N$ rapidly increases,
reaching a value very close to the final one. From Table 1, one
can see that its value depends quite strongly on the choice of
$Q$ and $\lambda$: $N$ grows if $Q$ increases and/or $\lambda$
is small. In particular, with $Q=5\div 10~M_{Pl}$ $N$ is of the
order $10^{2}$, which is the value needed to solve the flatness
problem of standard cosmology.

When the system reaches the absolute minimum of $V_{eff}$, $\varphi_c$ starts
oscillating around $\varphi_{c,m}$. For all cases reported in Table 1, for
which the fermion coupling constant has been chosen to be quite large, $f=1$,
the reheating temperature, $T_{RH}$, evaluated when
$\rho_{\sigma,\varphi_c}=\rho_{R}$, is of the order of $10^{14} \div
10^{15}~GeV$. Interestingly this value is the one typical of GUT theory energy
scales. If the value of $f$ is decreased, the reheating, in the one-loop
approximation we have considered, becomes less efficient. In fact, already for
$f\sim .1$ the energy injection {}from the $\sigma-\varphi_c$ system does not
compensate the dilution of the radiation energy due to the universe expansion.
Therefore, the value of the ratio $\rho_R /\rho_{\sigma,\varphi_c}$ reaches a
maximum smaller than the unity, and then starts to decrease again. In these
conditions the universe would continue to evolve in a $\rho_{\sigma,\varphi_c}$
matter dominated regime. This also happens for one of the cases reported in
Table 1, namely for the choice $Q=1$ and $\lambda=0.1$. The corresponding value
for the $e$-fold number, which is however unsatisfactorily small, should be
understood as the one at the beginning of the matter dominated era.

\section{Conclusions and remarks}

This paper represents a first step towards a systematic study of the
cosmological implications of gauge theory and gravitation models inspired by
noncommutative geometry. We have investigated this issue following
\cite{ConnesLott,Zurich,Zurgrav}, where gravitation and gauge theories are
combined in the framework of noncommutative geometry. By virtue of the
two-sheeted nature of the manifold, different scalar fields appear due to
gravity and gauge connections. The scalar dynamics naturally provides a chaotic
inflationary scenario in which, just after the Planck epoch, an initial field
configuration ($\sigma(0),\varphi_c(0)$) originates from quantum fluctuations
with a corresponding energy density of the order of $M_{Pl}^4$. The evolution
{}from the above initial condition is then fixed by the theory. A remarkable
aspect of this scenario is that the link between the scalar fields, the
distance along the {\it discrete dimension}, and the role played by the gravity
in the doubled manifold $\overline{\cal M}$, naturally fixes the initial values
for $\sigma$ and $\varphi_c$. This contrasts with the arbitrariness of the
usual models.

In this first analysis we have chosen an abelian gauge theory coupled with
gravity, which represents the simplest model capturing all the essential
ingredients of the dynamics. As is well-known, in noncommutative geometry to
have a nontrivial scalar interaction (scalar potential) more than one fermion
generation is needed.

We have considered the one-loop corrections to the classical potential due to
the $\delta \varphi_i$ quantum fluctuations introducing a cut-off
regularization; differently than in \cite{Ringwald} a renormalization procedure
cannot be applied since the $\varphi_c$ vacuum expectation value explicitly
depends on $\sigma$, as explained in section 3.1. Since from our numerical
simulation it comes out that this field reaches its stationary point $\sigma_m$
quite soon, a similar procedure can be applied in fact for the subsequent
evolution in which only the $\varphi_c$ field is still varying while $\sigma$
remains frozen at $\sigma_m$. To study the inflationary phase, however, since
the main contribution to the $e$-fold number comes from the epoch in which
$\sigma$ is rapidly evolving, we were forced to use the one-loop potential
explicitly depending on the cut-off parameter $Q$, for which we have considered
values of the order of the Planck mass.

The results show that, for reasonable values of the parameters, the value of
$N$ is of the right order of magnitude ($N \sim 100$) and the distance $d_5$ of
the two sheets of the manifold $\overline{\cal M}$ remains of the order of the
Planck length. It is interesting to note that it is the presence of the
discrete dimension which is responsible for the inflationary expansion; this
can be understood by realising that in the model it is $d_5^{-1}$ which play
the role of the effective cosmological constant. It is possible, with different
choices of the parameters, to have $d_5$ significantly growing as well, but
this can only be done at the price of a much smaller $e$-fold.

The reheating phase has been studied by considering the contribution at one
loop to the $\langle \delta \varphi_i^2 \rangle$ of the interactions with
fermion pairs. This introduces a dissipative term into the $\varphi_c$ and
$\sigma$ equations of motion which describe, at this level of approximation,
the energy release of the scalar fields to the relativistic degrees of freedom.
We found that, provided the coupling constant is large enough, $f \sim 1$, a
complete and efficient reheating is obtained and that the corresponding
reheating temperature is of the order of $10^{14} \div 10^{15} GeV$, well below
the initial energy scale, which is set by the Planck scale. As we already
pointed out, the approximation we have used is unsuitable for a {\it detailed}
analysis of the reheating phenomenon. We can nevertheless conclude that there
are strong indications of the presence of an efficient reheating stage.

There are at least two subjects which would be interesting to study further.
First of all a more realistic model, based on a nonabelian gauge symmetry
group, should be analyzed in more details, to see its implications for the
inflationary scenarios. Incidentally the presence of couplings of the scalar
fields with the gauge bosons should increase, in this case, the efficiency of
the reheating mechanism.

It has been shown that in the model we have considered, the distance $d_5$ at
the end of the inflationary era remains of the order of the Planck length; it
remains an open question, however, if its subsequent evolution is non-trivial,
namely that when the universe continues to expand after inflation, its value
gradually increases, up to the inverse top quark mass, which is the value
needed in order to reproduce the electroweak standard model. It seems that the
idea of looking at $d_5$ as undergoing a dynamical evolution is quite
important, since this could produce the onset of several phase transitions
which took place during the expansion of the universe.

\newpage
\large
\bigskip\bigskip
\par\noindent
{\bf Table 1.}\\
\bigskip
\begin{tabular}{|c|c|c|c|}
\hline
 & & & \\
$Q$ & $\lambda$  & $N$ & $T_{RH}$ \\
$(M_{Pl})$ & & &  $(10^{15}~GeV)$\\
& & & \\
\hline
& & & \\
$10$ & $0.1$  & $130$ & $5.$  \\
& & &  \\
$5$ & $0.01$ & $110$ & $.6$  \\
& & & \\
$5$ & $0.1$ & $40$ & $2.$  \\
& & & \\
$1$ & $0.01$ & $20$ & $.1$  \\
& & & \\
$1$ & $0.1$ & $10$ & $\mbox{no reheat.}$\\
& & & \\
\hline
\end{tabular}
\newpage

\noindent
{\Large \bf Figure Captions}

\normalsize

\begin{itemize}
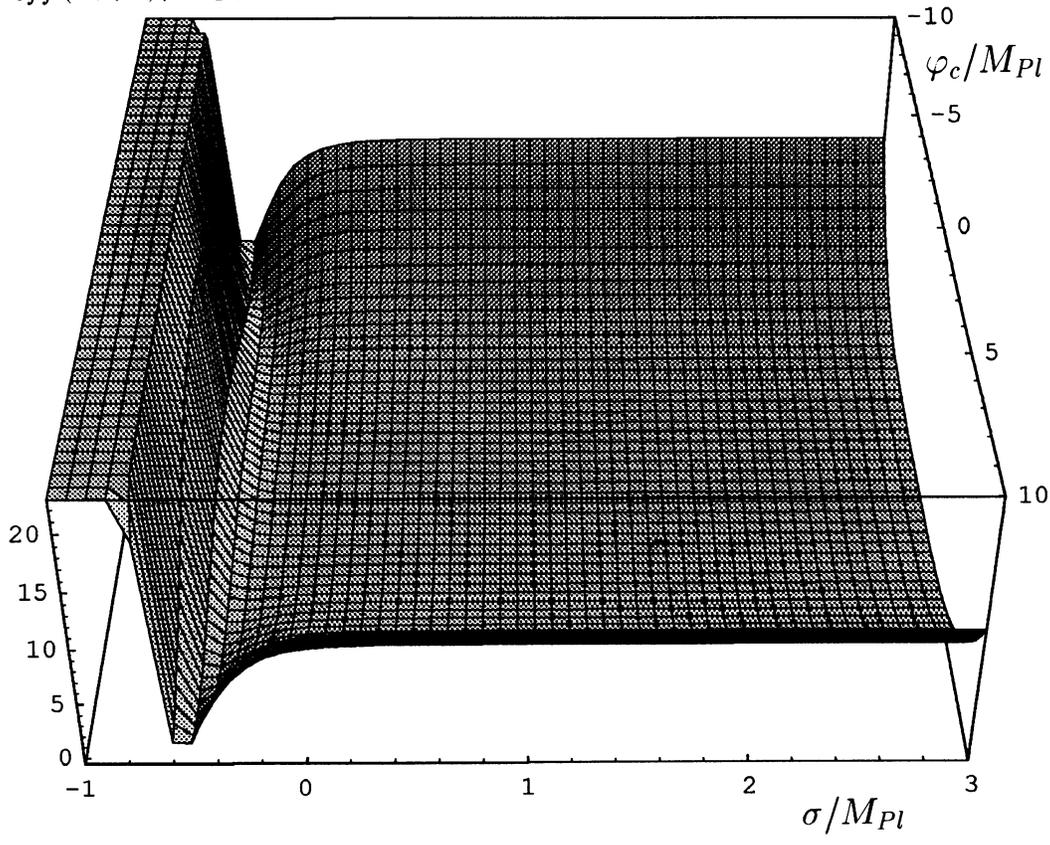
\item[Figure 1.] The effective potential $V_{eff}(\sigma,\varphi_c)$
is plotted for $\lambda=.1$ and $Q = 10 ~M_{Pl}$ and $m = 1~M_{Pl}$.
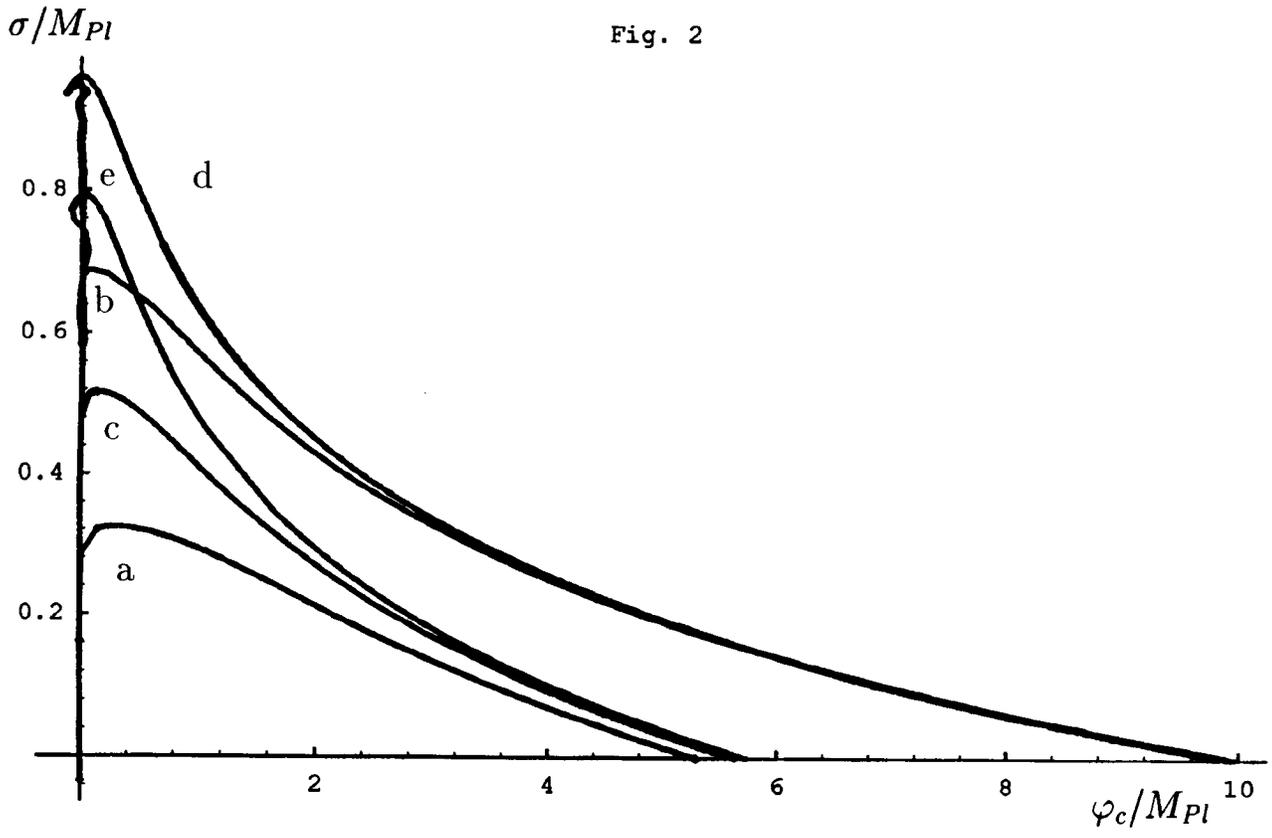
\item[Figure 2.]  The trajectories in the $\sigma-\varphi_c$ plane
corresponding to the values of the parameters reported in Table 1, are shown.
The labels refer to the order in which they are considered in the
Table, from top to bottom.
\end{itemize}

\vspace{3cm}

\noindent
{\Large \bf Table Caption}

\begin{itemize}
\item[Table 1.] The $e$-fold number, $N$, and the reheating temperature,
$T_{RH}$, are shown for some choices of the parameters $Q$, $\lambda$
(the values correspond to $f=1.$).
\end{itemize}
\end{document}